\documentclass[prl,twocolumn,tightenlines,superscriptaddress,nofootinbib,
showpacs]{revtex4}
\usepackage{amssymb,latexsym}
\usepackage{amsmath,amsbsy,bbm}
\usepackage{epsfig,bm}
\usepackage{graphicx,comment}
\usepackage{color}
\usepackage{soul}
\usepackage{csquotes}
\usepackage{placeins}
\usepackage[normalem]{ulem}
\begin{document}

\title{Validity of Harris criterion for two-dimensional quantum 
spin systems with quenched disorder}

\author{Jhao-Hong Peng}
\affiliation{Department of Physics, National Taiwan Normal University,
88, Sec.4, Ting-Chou Rd., Taipei 116, Taiwan}
\author{L.-W. Huang}
\affiliation{Department of Physics, National Taiwan Normal University,
88, Sec.4, Ting-Chou Rd., Taipei 116, Taiwan}
\author{D.-R. Tan}
\affiliation{Department of Physics, National Taiwan Normal University,
88, Sec.4, Ting-Chou Rd., Taipei 116, Taiwan}
\author{F.-J. Jiang}
\email[]{fjjiang@ntnu.edu.tw}
\affiliation{Department of Physics, National Taiwan Normal University,
88, Sec.4, Ting-Chou Rd., Taipei 116, Taiwan}
%\vspace{-2cm}

\begin{abstract}
Inspired by the recent results regarding whether the Harris criterion
is valid for quantum spin systems, we have simulated a two-dimensional 
spin-1/2 Heisenberg model on the square lattice with a specific kind of 
quenched disorder using the quantum Monte Carlo (QMC) calculations.  
In particular, the considered quenched disorder has a tunable parameter 
$0\le p \le 1$ which can be considered as a measure of randomness. 
Interestingly, when the magnitude of $p$ increases from 0 to 0.9, 
at the associated quantum phase transitions the 
numerical value of the correlation length exponent $\nu$ grows
from a number compatible with the $O(3)$ result 0.7112(5) to a number 
slightly greater than 1. In other words, by varying $p$, $\nu$ can reach 
an outcome between 0.7112(5) and 1 (or greater). Furthermore, 
among the studied values of $p$, all the associated $\nu$ 
violate the Harris criterion except the one corresponding to $p=0.9$.
Considering the form of the employed disorder here,
the above described scenario should remain true for other
randomness if it is based on the similar idea as the
one used in this study. This is indeed confirmed by our preliminary
results stemming from investigating another disorder distribution.    
         
\end{abstract}

\vskip-0.25cm
%\pacs{12.39.Fe, 75.10.Jm, 75.40.Mg, 75.50.Ee}

\maketitle

%\section{Introduction}
\vskip-0.1cm
Introduction ---
Studying the effects resulting from disorder has always been one of the major
topics in both theoretical and experimental physics 
\cite{Fis94,Vaj020,San02,Csa03,Lin03,Lin06,Laf06,Voj06,Voj10,Yao10,Car11,Yu12,Yu121,Voj13,Nvs14}. 
This is because
the presence of disorder such as impurities may lead to extraordinary 
properties and phases of materials. In particular, the appearance 
of these exotic characteristics are due to the mutual 
influence between the quantum fluctuations
and disorder. Understanding the relevance
of disorder at quantum phase transitions also continues to attract a lot of
attention. This is especially true considering the recent development
regarding under what conditions will the celebrated Harris criterion
be valid \cite{Har74,Cha86,Mot00,San02d,Vaj02,Skn04,Yu05,San06d}. 
In other words, 
it is not clearly at all that 
with what specific features of a disorder 
distribution will a new universality class emerge at the 
studied quantum phase transition.    

For a phase transition, there are three possible
scenarios when disorder is present. Here we will focus on those related to the
Harris criterion. The Harris criterion was originally derived for classical 
systems and its statement is as follows. For a $D$-dimensional classical 
system with disorder, the correlation
length exponent $\nu$ must satisfy the inequality $\nu \ge 2/D$. 
If the $\nu$ of a clean model does not fulfill this inequality,
then when disorder is introduced (into the clean model), a new
universality class should obtain so that the described inequality
is realized, assuming the phase transition remains well defined.
Later the criterion was generalized to more generic situations
including certain quantum systems. 
We would like to emphasize the fact that for a $d$-dimensional 
quantum system with
quenched disorder, since the disorder is employed in the 
spatial dimension, the dimensionality $D$ appearing in the inequality
is $d$, not $d+1$ despite the quantum system can be mapped to a 
$d+1$-dimensional classical system.

While the validity of Harris criterion is
beyond doubt for classical models, the case of quantum spin systems is much 
more complicated. Particularly, at the moment only the outcome related to
the two-dimensional (2d) spin-1/2
Heisenberg model on a bilayer square lattice with bond dilution
satisfies the Harris criterion \cite{Vaj02,San02d,Skn04,Yu05,San06d}. 
Other kinds of quenched disorder,
including the configurational disorder considered in Ref. \cite{Yao10} as well 
as the one introduced in Ref.~\cite{Nvs14}, the resulting calculations always
indicate the Harris criterion is violated. Furthermore, the obtained
values of the correlation length exponent $\nu$ remain the 
same as that of their clean counterparts. To summarize, whether the celebrated
Harris criterion is valid for quantum spin systems is more involved 
than anticipated.  

Inspired by such an indecisive answer regarding the applicability of
Harris criterion for quantum spin systems, in this study we have carried
out a large scale quantum Monte Carlo (QMC) calculations for a two-dimensional
spin-1/2 Heisenberg model on the square lattice, starting from the
clean herringbone model and then introducing a specific kind of quenched
disorder into the clean system. In particular, the employed randomness 
distribution has a tunable
parameter $p$ (Which can take values from 0 to 1 and can be considered as 
a measure of randomness) so that one can 
investigate the impact of this parameter on the effectiveness of Harris 
criterion for the studied model. 

Remarkably, our QMC data indicate that as the magnitude of $p$ increases 
gradually from 0 to 0.9, the numerical value of $\nu$ grows from its O(3)
value 0.7112(5) \cite{Nig92,Cam02,Pel02,Wan06,Alb08,Wen09,Car10,Sac11} to a 
result slightly greater than 1. In other words, 
by varying $p$, the corresponding $\nu$ for the disordered systems studied in this
investigation can reach outcomes lie between 0.7112(5) and 1.  
Moreover, the $\nu$ resulting from the considered values of $p$
all violate the Harris criterion except the one related to $p=0.9$.
Our preliminary study of another disorder distribution following
the similar idea as that introduced above leads to the same conclusion, 
namely $\nu$ can take a value between 0.7112(5) and 1 as well. The results
demonstrated here indicate that a better understanding of the Harris criterion
from a theoretical point of view is on request.

\begin{figure}
\vskip-0.5cm
\begin{center}
\vbox{
\includegraphics[width=0.245\textwidth]{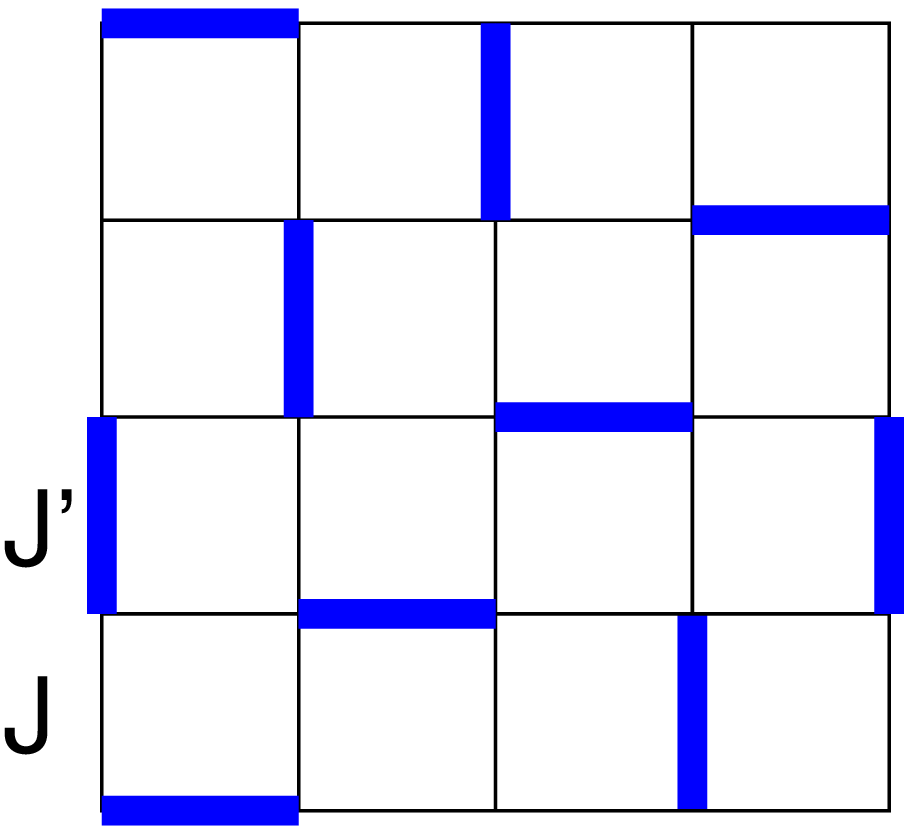}~~~
\includegraphics[width=0.22\textwidth]{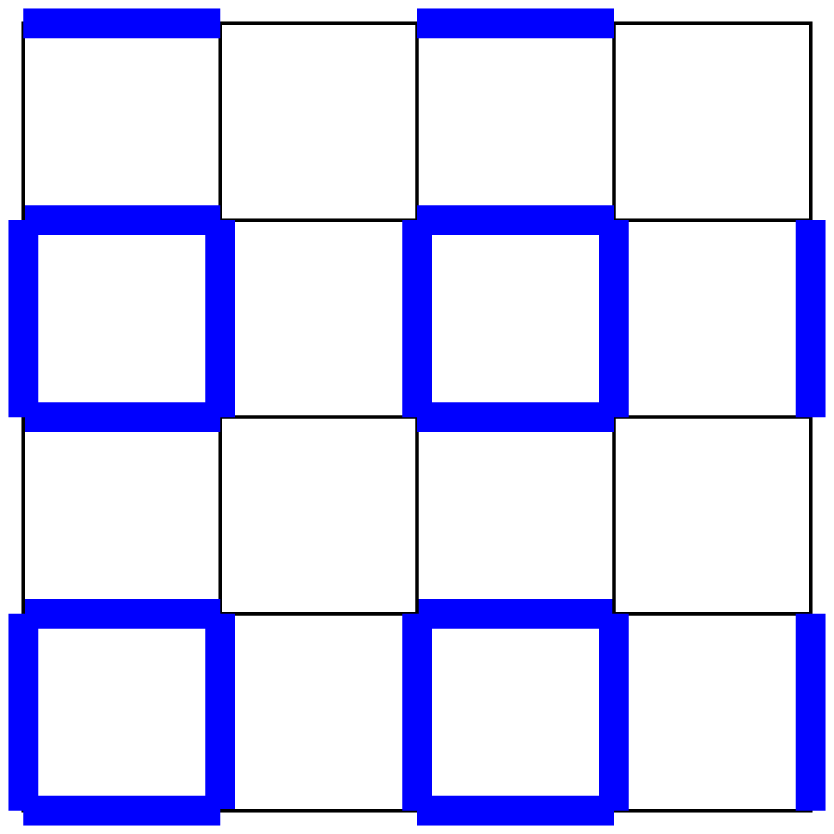}
}
\end{center}\vskip-0.5cm
\caption{The clean 2d herringbone (left) and plaquette (right) spin-1/2 
Heisenberg models on the square lattice investigated here. 
The antiferromagnetic 
coupling strengths for the thick and thin bonds are $J'$ and $J$, respectively.}
\label{model_fig1}
\end{figure}

%\section{Microscopic models and observables}
Microscopic models and observables ---
The Hamiltonian of the studied 2d (clean) herringbone spin-1/2 Heisenberg model 
on the square lattice is given by
\begin{eqnarray}
\label{hamilton}
H &=& \sum_{\langle ij \rangle}J_{ij}\,\vec S_i \cdot \vec S_{j} 
+ \sum_{\langle i'j' \rangle}J'_{i'j'}\,\vec S_{i'} \cdot \vec S_{j'},
\end{eqnarray}
where in Eq.~(1) $J_{ij}$ (which is set to 1 for any pair of ($i,j$) here) and 
$J'_{i'j'}$ are the antiferromagnetic 
couplings (bonds) connecting nearest neighbor spins $\langle  ij \rangle$ 
and $\langle  i'j' \rangle$ located at sites of the considered underlying  
square lattice, respectively, and $\vec S_{i} $ is the spin-1/2 operator at site $i$.  
The left panel of figure~\ref{model_fig1} demonstrates the typical clean herringbone 
models studied in the literature. The quenched disorder introduced into the (clean) 
system is based on the one employed in Ref.~\cite{Nvs14}. Specifically, for every bold bond in the left panel of fig.~\ref{model_fig1}, its
antiferromagnetic strength $J'$ takes the value of $1+(g-1)(1+p)$ 
or $1+(g-1)(1-p)$ with equal probability. Here $g>1$ and $0 \le p \le 1$.
As pointed out in Ref.~\cite{Nvs14}, the average and difference of $J'$ 
for these two types of bold bonds are given by $g$ and $2p(g-1)$,
respectively. Moreover, $p$ can be considered as a measure of disorder of the 
studied system. In our study, several values of $p$ are chosen and for each of
them, the corresponding phase transition is induced by tuning $g:= J'$.

To carry out the proposed investigation, particularly to examine the
validity of Harris criterion for the considered disordered model,
the observables 
first Binder ratio $Q_1$ and second Binder ratio $Q_2$ \cite{Bin81}, 
which are defined by
\begin{equation}
Q_1 = \frac{\langle |m_s^z| \rangle^2 }{\langle (m_s^z)^2\rangle} 
\end{equation}
and
\begin{equation}
Q_2 = \frac{\langle (m_s^z)^2 \rangle^2}{\langle (m_s^z)^4\rangle},
\end{equation} 
respectively, are measured in our calculations. 
The staggered magnetization density 
$m_s^z$ on a square lattice with linear box size $L$ appearing above is 
given by $m_s^z = \frac{1}{L^2}\sum_{i}(-1)^{i_1+i_2}S^z_i$ 
with $S^z_i$ being the third component of the spin-1/2 operator $\vec S_i$ at 
site $i$. Observable such as the one associated with spin 
stiffness $\rho_s$ 
has the following finite-size scaling expression close to the critical region 
\cite{Pel02,Wan06}
\begin{eqnarray}
\rho_s L^{z} &=& (1+b_0 L^{-\omega})f(tL^{1/\nu}), \nonumber \\
t &=& \frac{g-g_c}{g_c},
\end{eqnarray}
where $\nu$ and $\omega$ are the correlation length and the confluent exponents, respectively,
$z$ is the dynamic critical exponent, $b_0$ is some constant, and $f$ is a smooth function of its 
argument $tL^{1/\nu}$. To apply finite-size scaling for observable 
associated with $\rho_s$, one firstly
needs to determine $z$. Due to this, 
$Q_1$ and $Q_2$ are chosen as the relevant physical quantities for our 
investigation because their expected finite-size scaling formulas \cite{Wan06}
\begin{eqnarray}
Q_i &=& (1+b_{i} L^{-\omega})f_i(tL^{1/\nu}),\,i\in\{1,2\} \nonumber \\
t &=& \frac{g-g_c}{g_c},
\end{eqnarray}
do not contain the dynamic critical exponent $z$. Such a strategy, namely 
using $Q_1$ and $Q_2$ in our study, dramatically eliminates the computational complexity.

\begin{figure}%[ht!]
\vskip0.5cm
\begin{center}
\includegraphics[width=0.425\textwidth]{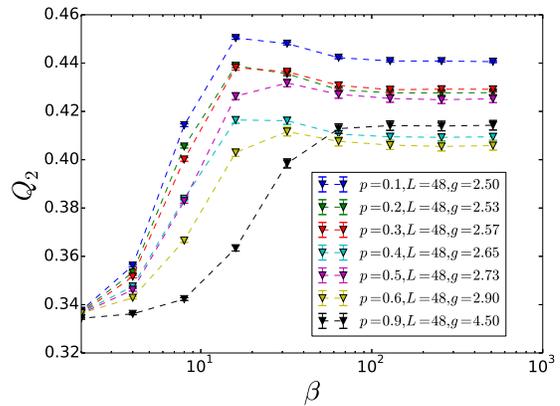}
\end{center}
\caption{Convergence in $\beta$ for several values of $p$ and $g$
using the $\beta$-doubling scheme. The box-sizes $L$ are 48 
for all the data presented in the figure.}
\label{fig1.5}
\end{figure}

\begin{table}[ht!]
%\label{tab1}
\begin{center}
\begin{tabular}{ccc}
\hline
$P$  & $\nu$ & $g_c$\\
\hline
\hline
0.0 & 0.703(5)& 2.4981(2) \\
\hline
0.1   & 0.701(5) & 2.5056(2) \\
\hline
0.2    & 0.724(5)& 2.5310(3) \\
\hline
0.3  & 0.754(6) & 2.5734(7) \\
\hline
0.4  & 0.781(7) & 2.6388(11) \\
\hline
0.5  & 0.816(9) & 2.7401(7) \\
\hline
0.6  & 0.843(11) & 2.8945(12) \\
\hline
0.9  & 1.04(3) & 4.828(18) \\
\hline
\end{tabular}
\end{center}
%\label{tab1}
\caption{Results of $\nu$ and $g_c$ obtained from the Bayesian analysis 
(and resampling). For each considered $p$, the listed outcome is based 
on the means and standatd deviations of all the results associated with it.
}
\label{tab0}
\end{table}

The numerical results ---
%\section{The numerical results}
For each of the studied $p$, to investigate the $g$ dependence of the correlation 
length exponent $\nu$ associated with it, we have carried out a large-scale QMC simulation 
using the stochastic series expansion (SSE) algorithm with very efficient 
operator-loop update \cite{San99,San10}. Furthermore, to obtain ground state properties
in an efficient manner, the $\beta$-doubling scheme described in Ref.~\cite{San02} is used 
in our simulations. Several hundred disordered
configurations, each with its own random seed, are generated for every considered set of parameters. 
It is also important
to notice that potentially there are two kinds of uncertainties for the used
observables, namely the one from Monte Carlo (MC) simulations and the one from
disorder averaging. We have carried out many trial simulations and have reached the conclusion that
with the MC sweeps employed in this study, the resulting errors of the considered quantities 
are indeed dominated by the disordered sample-to-sample fluctuation.

%\vskip1cm
\begin{figure}%[ht!]
\vskip0.5cm
\begin{center}
\vbox{
\includegraphics[width=0.425\textwidth]{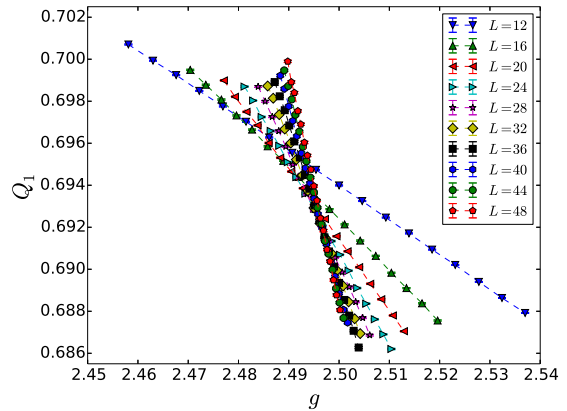}\vskip1cm
\includegraphics[width=0.425\textwidth]{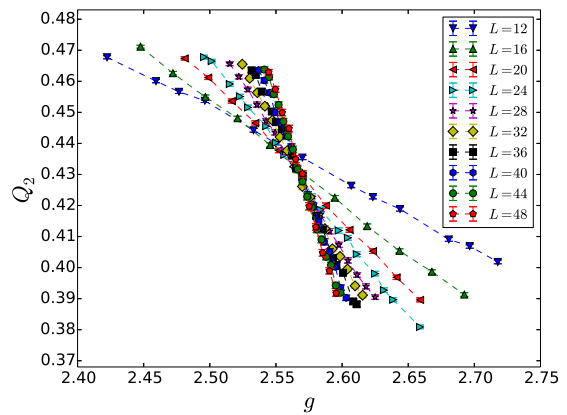}
}
\end{center}
\caption{$Q_1$ (top panel, $p=0$) and $Q_2$ (bottom panel, $p=0.3$) 
of various $L$ as functions of $g$
for the considered Herringbone models studied here. 
The dashed lines are added to guide the eye.}
\label{fig2}
\end{figure}

As already mentioned in the previous section, we have
focused on the observables $Q_1$ and $Q_2$. Several data sets regarding
convergence in $\beta$, as well as the final ground states values of $Q_1$ and
$Q_2$ are presented in Figs.~\ref{fig1.5} and ~\ref{fig2}. More data
of $Q_1$ and $Q_2$ are shown in the supplmental material.

The method used for the determination of $\nu$ (and $g_c$ as well) is the 
Bayesian analysis which is a rigorous mathematical approach. It is demonstrated
in Refs.~\cite{Har11,Har15} that the critical exponents calculated using the 
Bayesian 
analysis agree quantitatively with those determined by the conventional fits using
the idea of finite-size scaling. Here we follow the methods outlined in 
Ref.~\cite{For13}.
Moreover,
We have carried out many trial computations and
have arrived at the same conclusion as those in Refs. \cite{Har11,Har15},
namely the results obtained from the Bayesian 
analysis and the conventional finite-size scaling fits are consistent with
each other quantitatively.

The model
considered for the Bayesian analysis is the expected finite-size scaling
equations for $Q_1$ and $Q_2$ at a second order phase transition. Specifically,
the explicit expression of the model for the Bayesian analysis is given by
\begin{eqnarray}
(1+a_0L^{-\omega})(a_1 + a_2tL^{1/\nu} +a_3(tL^{1/\nu})^2+...).
\end{eqnarray}
Here $a_i$ for $i=0,1,2,...$ are some constants and $t = \frac{g-g_c}{g_c}$.
Moreover, this ansatz with up to third, fourth 
and fifth order in $tL^{1/\nu}$ are employed in the calculations of estimating the desired
physical quantities $\nu$ and $g_c$ with the data
of $Q_1$ and $Q_2$. Some constraints, such as the range of $g$ considered 
and the values of $\omega$ obtained, are taken in account in the 
procedure of analysis as well. 

Table \ref{tab0} summarizes the final quoted values of $\nu$ and $g_c$ for all 
the considered $p$. This table is based on the results of each individual $p$ 
obtained from the Bayesian analysis. Detailed outcomes of some values of $p$ 
are listed in the supplemental material.

For the clean model, the averaged $g_c$ and $\nu$ 
are given by 2.4981(2) and 0.703(5), respectively. The calculated $g_c$ is 
in nice agreement with 
the known results in the literature \cite{Fri11}. The determined $\nu$
for $p = 0$ is slightly smaller in magnitude than the expected
$O(3)$ value 0.7112(5). The largest $L$ used in the simulations conducted here is $L=48$. As
a result, the small deviation between 0.703(5) found here and 0.7112(5) can 
be easily accounted for by the cubic term introduced in Ref. \cite{Fri11} 
which will lead to anomalous large finite-size correction.

Table \ref{tab0} also implies that $g_c$ grows with $p$.
What's the most remarkable outcome shown in table \ref{tab0} is that, as the magnitude
of $p$ rises, the corresponding $\nu$ calculated
increases in size gradually from that of $p=0$ as well. Particularly
for $p \ge 0.3$ and $p \le 0.6$, the obtained $\nu$ from Bayesian analysis are all
statistically larger than 0.7112(5), but smaller than 1.0. In addition, 
for $p=0.9$, the associated $\nu$ is
around 1.0 with which the Harris criterion $\nu \ge 2/d$ is satisfied. 
To summarize,
the outcomes of our investigation, as shown in table \ref{tab0}, indicate that
for each employed $p$ such that $0.3 \le p < 0.9$, the resulting associated 
correlation length exponent neither stays as the $O(3)$ value $\nu=0.7112(5)$ 
nor satisfies the Harris criterion $\nu \ge 2/d$ = $1$. 
Moreover, for $p=0.9$, the Harris criterion is fulfilled. 
From the considered quantum spin system, we arrive at a scenario regarding 
the connection between $\nu$ and quenched disorder with a tunable randomness
strength not known before in the literature.  

It should be pointed out that the $\nu$ determined from $Q_1$ differ from that
related to $Q_2$ slightly. We attribute this to corrections not taken into account in the
analysis. Despite this, it is without doubt that both $Q_1$ and $Q_2$ will lead
to the scenario described above.  

Based on the explicit expression of the disorder taken into account here,
one expects that the obtained scenario should still be valid for other
randomness distributions using the similar idea as the one investigated above.
Motivated by this intuitive thought, apart from simulating the disordered 
system introduced previously, we have considered a quenched disorder for
the plaquette model (The right hand side panel of fig.~1). Specifically,
each of the bold bonds takes the antiferrognetic strength of $(1+K)J_c$ and 
$(1-K)J_c$ with probability $P$ and $1-P$, respectively. Here $0<P<1$ and we 
have used $K=0.5$. In addition, the $J_c$ appearing above is given
by 1.8230 which is the critical point of the clean plaquette model. With
such a set up, $P$ is the tunable variable for this model. The resulting 
$Q_1$ and $Q_2$ of this new model with the new
type of quenched disorder are shown in the supplemental materail.
Moreover, by applying typical fits with the conventional finite-size scaling equations 
to $Q_1$ and $Q_2$, we arrive at $\nu = 0.79(2)$ (and the critical point $P_c = 0.5520(16)$).   
This number $\nu = 0.79(2)$ is without doubt statistically different from both 
0.7112(5) and 1. By changing $D$ continuously, it is anticipated
that the corresponding $\nu$ will vary in a gradual manner. This is of high
similarity to the scenario associated with the Herringbone model
found earlier in this study.

Discussions and Conclusions ---
%\section{Discussions and Conclusions}
According to Ref.~\cite{Paz97}, based on the procedure of scaling employed
in this study, the Harris criterion should be 
$\nu_{\text{FS}} \ge 2/d$, where the subscript FS stands for finite size.
In particular, Ref.~\cite{Paz97} also demonstrates that the bulk correlation length 
$\nu$ can be obtained by a modified procedure and may violate the Harris 
criterion. Considering the facts that most of the outcomes obtained 
in this study violate $\nu_{\text{FS}} \ge 2/d$ and the variation among 
all the $g_c(p)$ determined here is not small, the observed scenario cannot be 
easily accounted for by the arguments in Ref.~\cite{Paz97}.
Results of some studies such as Refs.~\cite{Yao10,Kis97}
imply that the values of $\nu$ (or $\nu_{\text{FS}}$) calculated
do not depend on the disorder strength.
The scenario found in this study clearly is different from this and other 
established ones in the literature.

We would like to re-emphasize the following points. First of all, 
the $\nu_{\text{FS}}$ obtained here for $p \ge 0.3$ all violate
the Harris criterion $\nu_{\text{FS}} \ge 2/d$ except for $p=0.9$.  
Secondly, for any given $p$, the phase transition is due to dimerization,
namely two nearest neighboring spins (on the lattice) form a singlet.
Hence, theoretically it is anticipated that for two close by values 
of $p$, say $p=0.3$ and $p=0.4$, the calculated results of $\nu$
should be close to each other or even consistent within statistical error. 
As can be seen from table \ref{tab0}, this is not the case. In particular,
the difference of the $g_c$ between $p=0.0$ and $p=0.3$ is only around 3 
percent, yet a new critical exponent $\nu$ emerges for $p=0.3$.
It is interesting as well to notice that while in \cite{Yao10} 
the $g_c$ of the random plaquette model varies from that of
its clean counterpart by four percent, both models have the same
$O(3)$ exponent $\nu$ = 0.7112(5). Therefore, closeness the critical
point of a disordered system from that of it clean analogue, which
may be interpreted as the statement \enquote{ remains well defined } 
in the Harris criterion, is not crucial for the appearance of a new 
universality class.

Interestingly, based on the outcomes found here, it is likely that
for the considered model with the designed quenched disorder, the largest value of 
$\nu$ one can obtain should be around 1.1 to 1.2. This
number agrees with the results calculated by simulating quantum spin model on the bilayer
lattice with bond dilution \cite{San02d,Vaj02}. It is plausible that for a
disordered spin-1/2 system, whenever the associated $\nu$ fulfills the
Harris criterion, its value is in the range of 1 to 1.2. 

In conclusion, in order to obtain a theoretical explanation for the exotic scenario
observed in this study, a detailed exploration of the relevant theory than what's been accomplished 
for the Harris criterion is required.

%\vskip3cm

\vskip0.5cm
%\section{Acknowledgments}
%\vskip-0.5cm
This study is partially supported by MOST of Taiwan.

~
~
~

~
~
~
~
~
~
~
~
~
~
~
~
~

%\end{document}

%\documentclass[prl,twocolumn,tightenlines,superscriptaddress,nofootinbib]{revte%x4}
%\usepackage{amssymb,latexsym}
%\usepackage{amsmath,amsbsy,bbm}
%\usepackage{epsfig,bm}
%\usepackage{graphicx,comment}
%\usepackage{color}
%\usepackage{soul}
%\usepackage{csquotes}
%\usepackage{authblk}
%\usepackage[normalem]{ulem}
%\unitlength=1mm
 
%\begin{document} 

\section{Supplemental Material to " Validity of Harris criterion for 
two-dimensional quantum spin systems with quenched disorder "}

%\author{J.-H. Peng}
%\affiliation{Department of Physics, National Taiwan Normal University,
%88, Sec.4, Ting-Chou Rd., Taipei 116, Taiwan}
%\author{L.-W. Huang}
%\affiliation{Department of Physics, National Taiwan Normal University,
%88, Sec.4, Ting-Chou Rd., Taipei 116, Taiwan}
%\author{D.-R. Tan}
%\affiliation{Department of Physics, National Taiwan Normal University,
%88, Sec.4, Ting-Chou Rd., Taipei 116, Taiwan}
%\author{F.-J. Jiang}
%%\email[]{fjjiang@ntnu.edu.tw}
%\affiliation{Department of Physics, National Taiwan Normal University,
%88, Sec.4, Ting-Chou Rd., Taipei 116, Taiwan}
%%\vspace{-2cm}

%\maketitle

\subsection{Details of the results from Bayesian analysis for 
the Herringbone model}
In this section, we list the values of $\nu$
and $g_c$ obtained from analysis with various conditions for some of the 
considered $p$. Specially, we will brief describe
how these outcomes are obtained from the relevant data. 
The method used in obtaining the corresponding 
results of $\nu$ and $g_c$ from the data of $Q_1$ and $Q_2$ is the Bayesian 
analysis. In particular, the initial values of
$\nu$ and $g_c$ determined from the analysis are the {\it Maximum a posterior}
estimations (Which is defined as the result of having the least value of 
$\chi^2/{\text{DOF}}$ in this study due to the flat prior). In addition, the associated 
uncertainties are the 
standard deviations of the the resulting probability distributions. For each
$p$ and a fixed $L_m$ ($L_m$ is the smallest box size used in the analysis), 
several examinations are performed using Bayesian method with various range of 
$g$. Moreover, resampling with respect to these outcomes obtained by 
considering various range of $g$ are conducted, using the related uncertainties 
as the Gaussian noises. The $\nu$ and $g_c$ finally presented in the following tables
and used to calculate the outcomes of table 1 in the main text, are the means 
and standard deviations of the results calculated from the resampling 
procedures described above. 

\FloatBarrier
\begin{table}[!ht]
%\label{tab1}
\begin{center}
\scalebox{0.95}{
\begin{tabular}{ccccccc}
\hline
$P$  & $L_m$ & $tL^{1/\nu}$ & $\nu$&$\nu$ &$g_c$ & $g_c$\\
\hline
\hline
0.0 & 8& 3 & 0.709(1) &0.698(1) &2.4985(1)&2.4980(1)\\
\hline
0.0   & 8 & 4 & 0.709(1)&0.699(1) &2.4985(1)&2.4980(1)\\
\hline
0.0    & 8& 5 & 0.709(1)&0.699(1) &2.4985(1)&2.4980(1)\\
\hline
0.0  & 12 & 3& 0.707(1)&0.699(1)&2.4982(1)&2.4980(1)\\
\hline
0.0  & 12 & 4&0.707(1)&0.699(1)&2.4982(1)&2.4981(1)\\
\hline
0.0  & 12 & 5&0.707(1)&0.699(1)&2.4982(1)&2.4981(1)\\
\hline
0.0  & 16 & 3& 0.706(1)&0.700(1)&2.4980(1)&2.4980(1)\\
\hline
0.0  & 16 & 4&0.706(1)&0.700(1)&2.4980(1)&2.4980(1)\\
\hline
0.0  & 16 & 5& 0.707(1)&0.700(1)&2.4980(1)&2.4980(1)\\
\hline
\end{tabular}}
\end{center}
%\label{tab1}
%\vskip-0.5cm
\caption{Results of $\nu$ and $g_c$ obtained from Bayesian analysis (and resampling) for the clean Herringbone model.
The second and third columns are the minimum box size and the 
order of polynomials in $tL^{1/\nu}$ (including subleading correction) 
used in the analysis, respectively. The $\nu$ and $g_c$ shown in the 4th (5th) and 6th (7th) 
columns are determined from $Q_1$ ($Q_2$). }
\label{tab1}
\end{table}
\FloatBarrier
\FloatBarrier
\begin{table}[!ht]
%\label{tab3}
\begin{center}
\scalebox{0.95}{
\begin{tabular}{ccccccc}
\hline
$P$  & $L_m$ & $tL^{1/\nu}$ & $\nu$&$\nu$ &$g_c$ & $g_c$\\
\hline
\hline
0.2 & 8& 3 & 0.730(3) &0.721(3) &2.5314(3)&2.5306(2)\\
\hline
0.2   & 8 & 4 & 0.730(3)&0.721(3) &2.5313(3)&2.5306(2)\\
\hline
0.2    & 8& 5 & 0.730(3)&0.721(3) &2.5314(3)&2.5305(2)\\
\hline
0.2  & 12 & 3& 0.727(3)&0.719(3)&2.5310(4)&2.5309(3)\\
\hline
0.2  & 12 & 4&0.728(3)&0.719(3)&2.5311(4)&2.5310(3)\\
\hline
0.2  & 12 & 5&0.728(3)&0.719(3)&2.5309(4)&2.5310(3)\\
\hline
0.2  & 16 & 3& 0.727(3)&0.719(3)&2.5309(4)&2.5312(4)\\
\hline
0.2  & 16 & 4&0.726(3)&0.718(3)&2.5309(4)&2.5311(4)\\
\hline
0.2  & 16 & 5& 0.727(3)&0.719(3)&2.5310(4)&2.5310(4)\\
\hline
\end{tabular}}
\end{center}
%\vskip-0.5cm
\caption{Results of $\nu$ and $g_c$ obtained from Bayesian analysis 
(and resampling) for the disordered Herringbone model with $p=0.2$.
The second and third columns are the minimum box size and the
order of polynomials in $tL^{1/\nu}$ (including subleading correction) 
used in the analysis, respectively. The $\nu$ and $g_c$ shown in the 4th (5th) and 6th (7th) 
columns are determined from $Q_1$ ($Q_2$).}
\label{tab3}
\end{table}
\FloatBarrier
%\vskip-0.5cm
\FloatBarrier
\begin{table}[!ht]
%\label{tab4}
\begin{center}
\scalebox{0.95}{
\begin{tabular}{ccccccc}
\hline
$P$  & $L_m$ & $tL^{1/\nu}$ & $\nu$&$\nu$ &$g_c$ & $g_c$\\
\hline
\hline
0.3 & 8& 3 & 0.757(6) &0.745(6) &2.5744(7)&2.5738(5)\\
\hline
0.3   & 8 & 4 & 0.758(6)&0.748(6) &2.5744(7)&2.5739(5)\\
\hline
0.3    & 8& 5 & 0.756(6)&0.745(6) &2.5741(7)&2.5737(5)\\
\hline
0.3  & 12 & 3& 0.759(6)&0.748(6)&2.5729(6)&2.5730(5)\\
\hline
0.3  & 12 & 4&0.759(6)&0.750(6)&2.5726(5)&2.5729(5)\\
\hline
0.3  & 12 & 5&0.762(6)&0.751(6)&2.5725(5)&2.5727(5)\\
\hline
0.3  & 16 & 3& 0.760(6)&0.751(6)&2.5735(5)&2.5739(6)\\
\hline
0.3  & 16 & 4&0.763(6)&0.751(6)&2.5730(5)&2.5739(6)\\
\hline
0.3  & 16 & 5& 0.761(6)&0.752(6)&2.5726(5)&2.5739(6)\\
\hline
\end{tabular}}
\end{center}
%\vskip-0.5cm
\caption{Results of $\nu$ and $g_c$ obtained from Bayesian analysis 
(and resampling) for the disordered Herringbone model with $p=0.3$.
The second and third columns are the minimum box size and the
order of polynomials in $tL^{1/\nu}$ (including subleading correction) 
used in the analysis, respectively. The $\nu$ and $g_c$ shown in the 4th (5th) and 6th (7th) 
columns are determined from $Q_1$ ($Q_2$).}
\label{tab4}
\end{table}
\FloatBarrier
%\vskip-0.5cm
\FloatBarrier
\begin{table}[!ht]
%\label{tab5}
\begin{center}
\scalebox{0.95}{
\begin{tabular}{ccccccc}
\hline
$P$  & $L_m$ & $tL^{1/\nu}$ & $\nu$&$\nu$ &$g_c$ & $g_c$\\
\hline
\hline
0.4 & 8& 3 & 0.787(5) &0.773(4) &2.6401(8)&2.6402(7)\\
\hline
0.4   & 8 & 4 & 0.785(5)&0.772(5) &2.6396(8)&2.6402(7)\\
\hline
0.4    & 8& 5 & 0.780(5)&0.767(5) &2.6403(8)&2.6404(7)\\
\hline
0.4  & 12 & 3& 0.790(5)&0.778(5)&2.6378(5)&2.6384(6)\\
\hline
0.4  & 12 & 4&0.789(5)&0.777(5)&2.6375(5)&2.6382(5)\\
\hline
0.4  & 12 & 5&0.783(5)&0.773(5)&2.6379(5)&2.6385(5)\\
\hline
0.4  & 16 & 3& 0.791(5)&0.780(5)&2.6379(5)&2.6386(6)\\
\hline
0.4  & 16 & 4&0.790(5)&0.780(5)&2.6378(6)&2.6386(6)\\
\hline
0.4  & 16 & 5& 0.784(5)&0.778(5)&2.6382(6)&2.6388(6)\\
\hline
\end{tabular}}
\end{center}
%\vskip-0.5cm
\caption{Results of $\nu$ and $g_c$ obtained from Bayesian analysis 
(and resampling) for the disordered Herringbone model with $p=0.4$.
The second and third columns are the minimum box size and the
order of polynomials in $tL^{1/\nu}$ (including subleading correction) 
used in the analysis, respectively. The $\nu$ and $g_c$ shown in the 4th (5th) and 6th (7th) 
columns are determined from $Q_1$ ($Q_2$).}
\label{tab5}
\end{table}
\FloatBarrier
%\vskip-0.5cm
\FloatBarrier
\begin{table}[!ht]
%\label{tab6}
\begin{center}
\scalebox{0.95}{
\begin{tabular}{ccccccc}
\hline
$P$  & $L_m$ & $tL^{1/\nu}$ & $\nu$&$\nu$ &$g_c$ & $g_c$\\
\hline
\hline
0.5 & 8& 3 & 0.833(7) &0.816(6) &2.7397(9)&2.7403(8)\\
\hline
0.5   & 8 & 4 & 0.824(7)&0.809(7) &2.7403(10)&2.7406(9)\\
\hline
0.5    & 8& 5 & 0.818(7)&0.806(7) &2.7400(10)&2.7408(9)\\
\hline
0.5  & 12 & 3& 0.832(7)&0.815(6)&2.7393(7)&2.7405(9)\\
\hline
0.5  & 12 & 4&0.818(7)&0.808(7)&2.7398(8)&2.7409(9)\\
\hline
0.5  & 12 & 5&0.818(7)&0.805(7)&2.7397(8)&2.7407(9)\\
\hline
0.5  & 16 & 3& 0.826(8)&0.811(7)&2.7389(8)&2.7406(9)\\
\hline
0.5  & 16 & 4&0.817(7)&0.805(7)&2.7398(8)&2.7410(9)\\
\hline
0.5  & 16 & 5& 0.814(8)&0.806(7)&2.7392(8)&2.7404(9)\\
\hline
\end{tabular}}
\end{center}
%\vskip-0.5cm
\caption{Results of $\nu$ and $g_c$ obtained from Bayesian analysis 
(and resampling) for the disordered Herringbone model with $p=0.5$.
The second and third columns are the minimum box size and the
order of polynomials in $tL^{1/\nu}$ (including subleading correction) 
used in the analysis, respectively. The $\nu$ and $g_c$ shown in the 4th (5th) and 6th (7th) 
columns are determined from $Q_1$ ($Q_2$).}
\label{tab6}
\end{table}
\FloatBarrier

\FloatBarrier
\begin{table}[!ht]
%\label{tab7}
\begin{center}
\scalebox{0.95}{
\begin{tabular}{ccccccc}
\hline
$P$  & $L_m$ & $tL^{1/\nu}$ & $\nu$&$\nu$ &$g_c$ & $g_c$\\
\hline
\hline
0.9 & 8& 3 & 1.086(9) &1.039(7) &4.854(14)&4.836(14)\\
\hline
0.9   & 8 & 4 & 1.090(10)&1.045(8) &4.839(15)&4.824(13)\\
\hline
0.9    & 8& 5 & 1.083(10)&1.043(8) &4.833(15)&4.822(14)\\
\hline
0.9  & 12 & 3& 1.042(9)&1.021(9)&4.823(14)&4.852(17)\\
\hline
0.9  & 12 & 4&1.042(9)&1.021(8)&4.807(11)&4.832(16)\\
\hline
0.9  & 12 & 5&1.040(9)&1.021(9)&4.812(14)&4.828(17)\\
\hline
0.9  & 16 & 3&1.023(10)&1.013(9)&4.803(13)&4.855(17)\\
\hline
0.9  & 16 & 4&1.022(9)&1.011(9)&4.802(13)&4.840(17)\\
\hline
0.9  & 16 & 5& 1.024(10)&1.010(9)&4.807(14)&4.841(19)\\
\hline
\end{tabular}}
\end{center}
%\vskip-0.5cm
\caption{Results of $\nu$ and $g_c$ obtained from Bayesian analysis 
(and resampling) for the disordered Herringbone model with $p=0.9$.
The second and third columns are the minimum box size and the
order of polynomials in $tL^{1/\nu}$ (including subleading correction)
used in the analysis, respectively. The $\nu$ and $g_c$ shown in the 4th (5th) and 6th (7th) 
columns are determined from $Q_1$ ($Q_2$).}
\label{tab8}
\end{table}
\FloatBarrier

\vskip-2cm

\subsection{More Data of $Q_1$ and $Q_2$}
In this section, more data for the disordered Herringbone 
and plaquette models are presented.  

\vskip-0.5cm
\FloatBarrier
\begin{figure}%[!ht]
\vskip-0.5cm
\begin{center}
\vbox{
\includegraphics[width=0.39\textwidth]{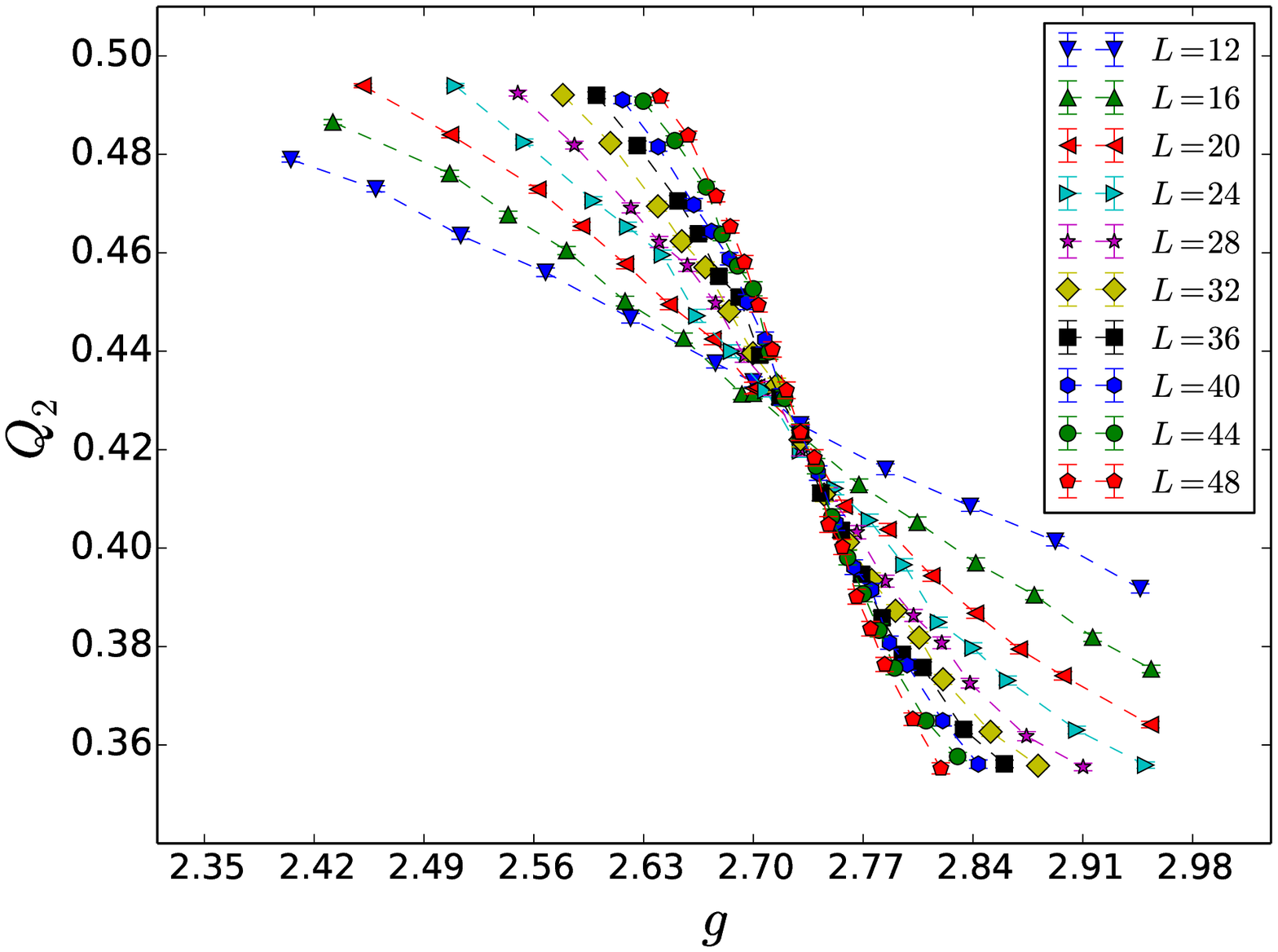}\vskip0.1cm
\includegraphics[width=0.39\textwidth]{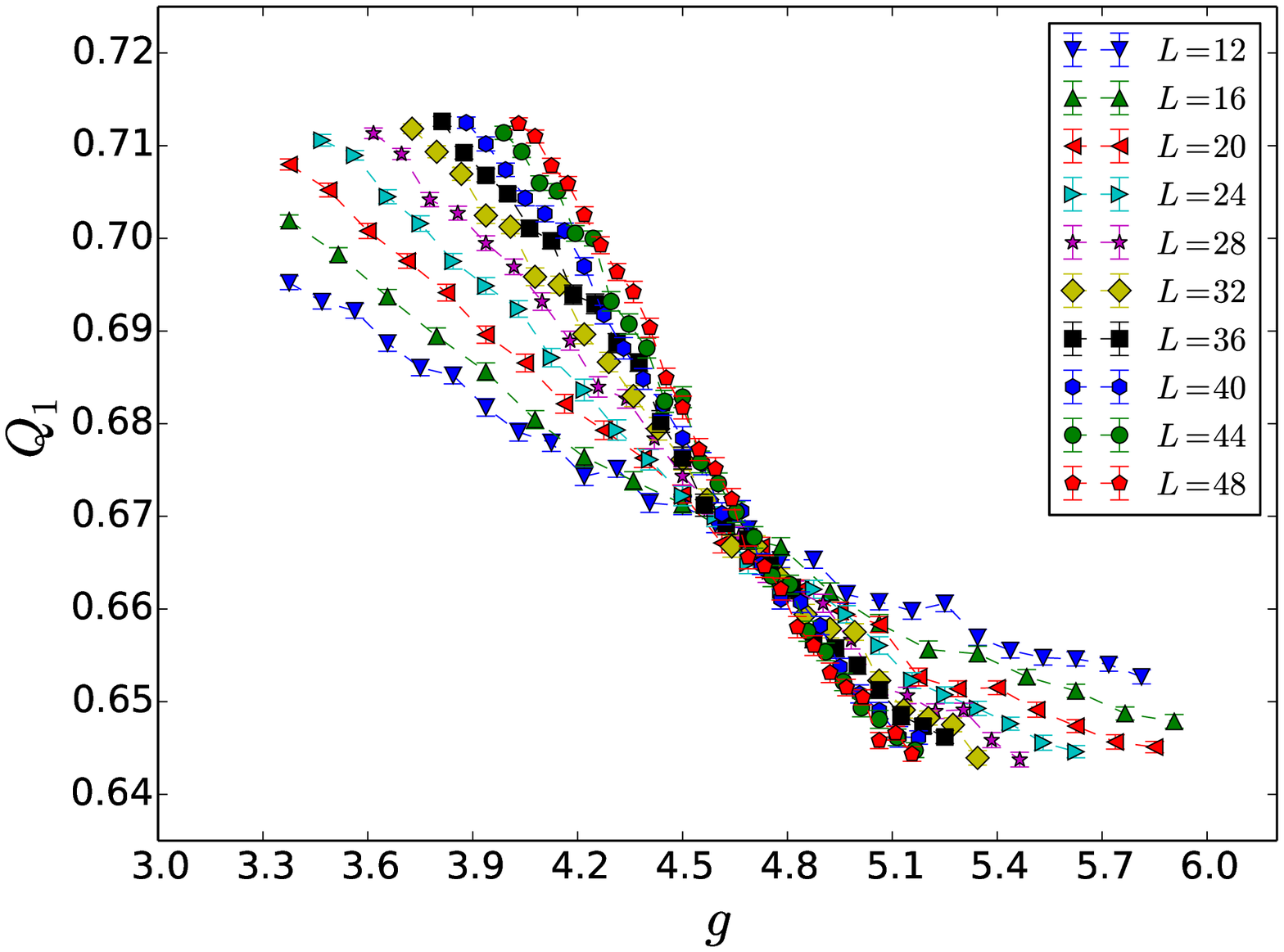}
}
\end{center}
\vskip-0.5cm
\caption{$Q_2$ (top panel, $p=0.5$) and $Q_1$ (bottom panel, $p=0.9$) 
of various $L$ as functions of $g$
for the considered disordered Herringbone models studied here. 
The dashed lines are added to guide the eye.}
\label{fig3}
\end{figure}
\FloatBarrier

\FloatBarrier
\begin{figure}[!ht]
\vskip-0.5cm
\begin{center}
\vbox{
\includegraphics[width=0.39\textwidth]{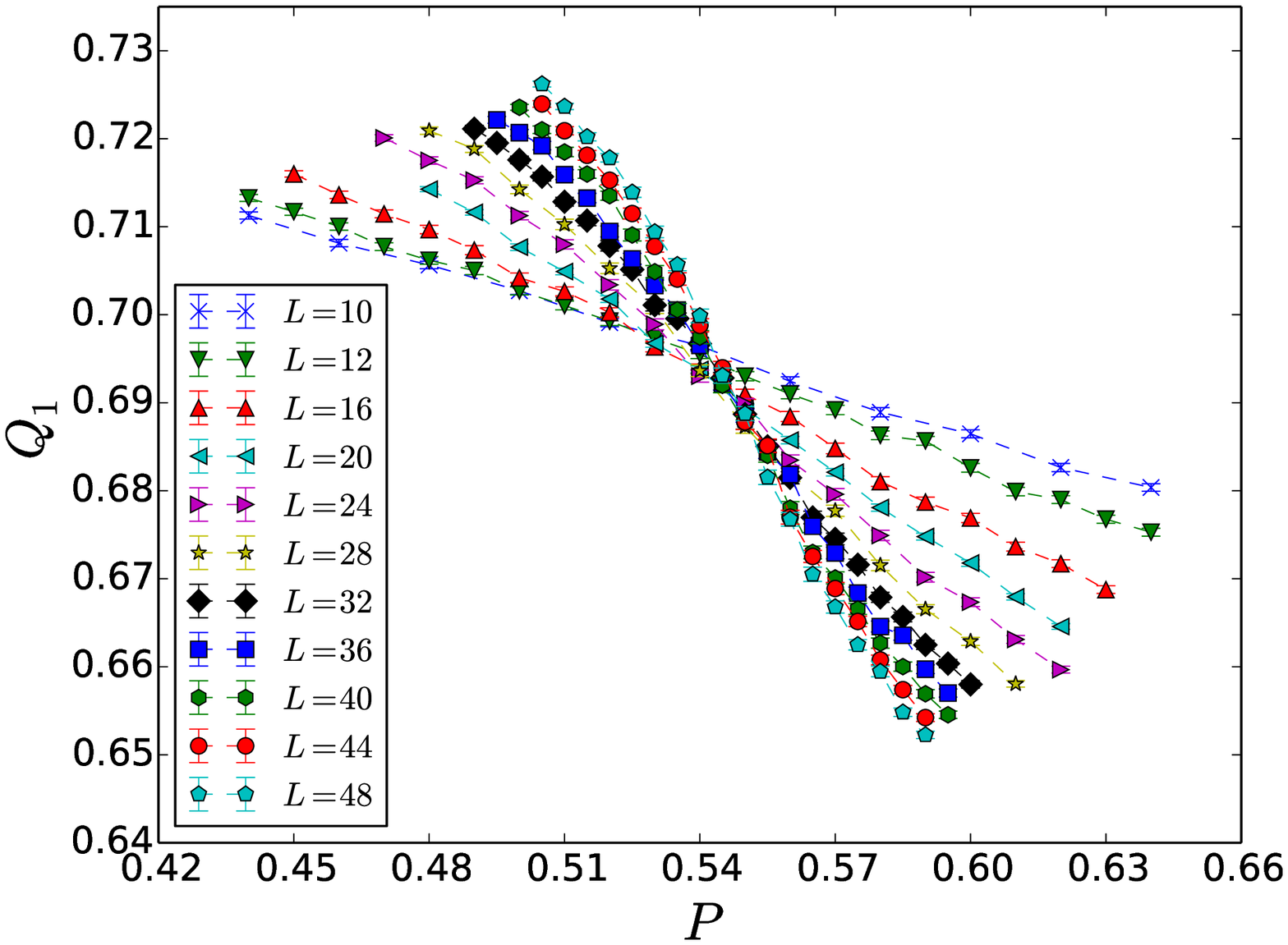}\vskip0.1cm
\includegraphics[width=0.39\textwidth]{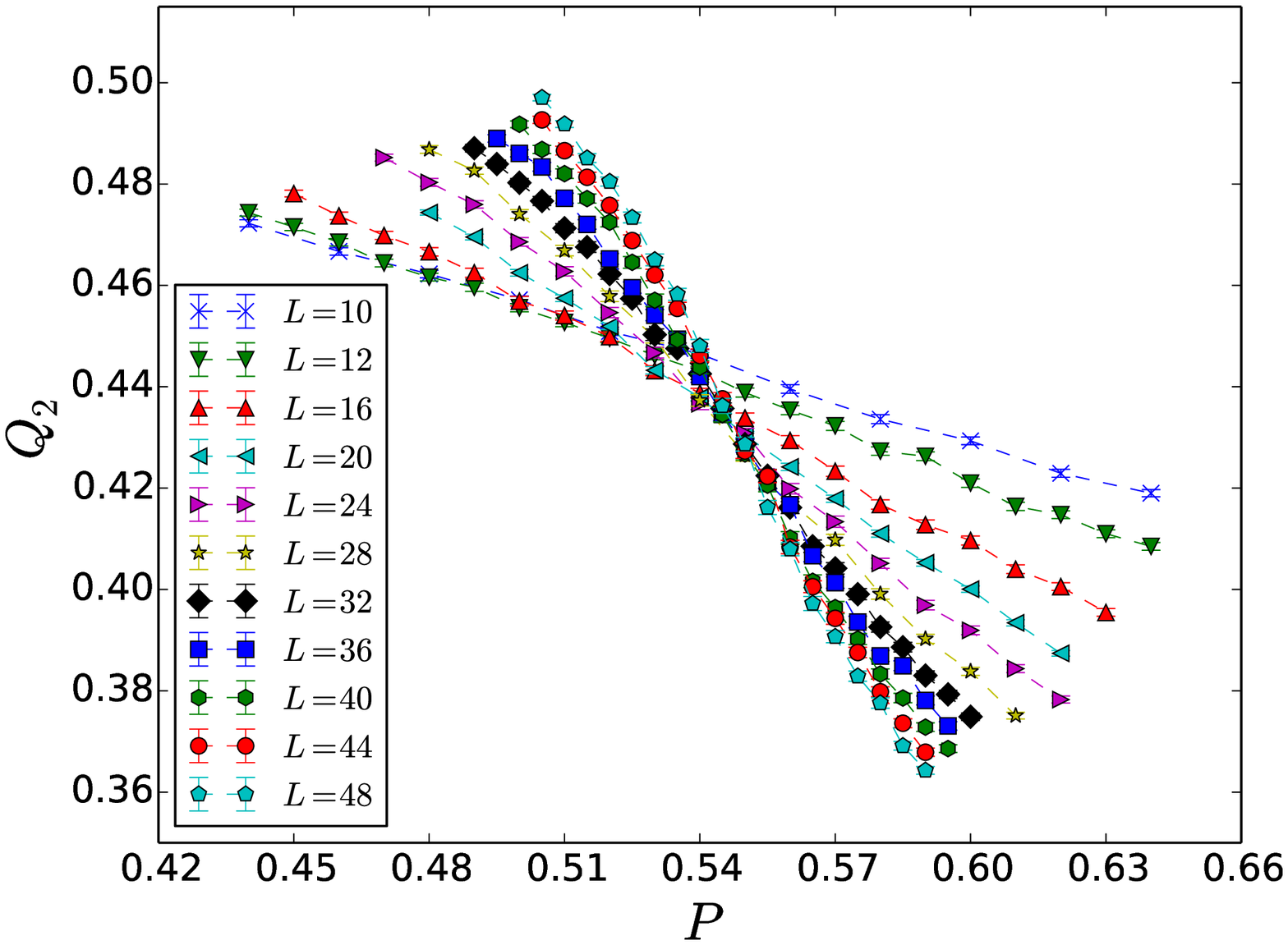}
}
\end{center}
\vskip-0.5cm
\caption{$Q_1$ (top panel) and $Q_2$ (bottom panel) of various $L$
as functions of $P$ for the considered disordered plaquette model studied 
here. The dashed lines are added to guide the eye.}
\label{fig5}
\end{figure}
\FloatBarrier


\begin{thebibliography}{99}

\bibitem{Fis94} D. S. Fisher, Phys. Rev. B {\bf 50}, 3799 (1994).

\bibitem{Vaj020}
O. Vajk, P. Mang, M. Greven, P. Gehring, and J. Lynn,
Science {\bf 295}, 1691 (2002).

\bibitem{San02}
A.~W.~Sandvik, Phys. Rev. B {\bf 66}, 024418 (2002).

\bibitem{Csa03} G. A. Cs$\acute{a}$thy, J. D. Reppy, and M. H. W. Chan, Phys.
Rev. Lett. {\bf 91}, 235301 (2003).

\bibitem{Lin03}
Y.-C. Lin, R. M$\acute{e}$lin, H. Rieger, and F. Igl$\acute{o}$i, 
Phys. Rev. B {\bf 68}, 024424 (2003).

\bibitem{Lin06} Y.-C. Lin, H. Rieger, N. Laflorencie, and F. Ig$\acute{o}$i, 
Phys. Rev. B {\bf 74}, 024427 (2006).

\bibitem{Laf06}
N. Laflorencie, S. Wessel, A. L$\ddot{a}$uchli, and H. Rieger, 
Phys. Rev. B {\bf 73}, 060403(R) (2006).


\bibitem{Voj06}
Thomas Vojta, J. Phys. A 39, R143-R205 (2006)

\bibitem{Voj10}
T. Vojta, J. Low Temp. Phys. 161, 299 (2010).

\bibitem{Yao10}
Dao-Xin Yao, Jonas Gustafsson, E.~W.~Carlson, and Anders W.~Sandvik,
Physical Review B, {\bf 82}, 172409 (2010).

\bibitem{Car11}
P. Carretta, G. Prando, S. Sanna, R. De Renzi, C. Decorse,
and P. Berthet, Phys. Rev. B {\bf 83}, 180411(R) (2011).

\bibitem{Yu12}
R. Yu, C. F. Miclea, F. Weickert, R. Movshovich, A.
Paduan-Filho, V. S. Zapf, and T. Roscilde, Phys. Rev. B
{\bf 86}, 134421 (2012).

\bibitem{Yu121} R. Yu, L. Yin, N. S. Sullivan, J. S. Xia, C. Huan, A.
Paduan-Filho, N. F. Oliveira Jr, S. Haas, A. Steppke, C.
F. Miclea, F. Weickert, R. Movshovich, E.-D. Mun, B. L.
Scott, V. S. Zapf, T. Roscilde, and A. Kitaev, Nature {\bf 489},
379 (2012).

\bibitem{Voj13}
Thomas Vojta, AIP Conference Proceedings 1550, 188 (2013).

\bibitem{Nvs14}
Nvsen Ma, Anders W. Sandvik, and Dao-Xin Yao, Phys. Rev. B {\bf 90}, 104425 (2014).

\bibitem{Har74}
A.~B.~Harris, J. Phys. C {\bf 7}, 1671 (1974).

\bibitem{Cha86}
J.~T.~Chayes, L.~Chayes, D.~S.~Fisher, and T.~Spencer, Phys. Rev.
Lett. {\bf 57}, 2999 (1986).

\bibitem{Mot00}
O. Motrunich, S.C. Mau, D.A. Huse, and D.S. Fisher,
Phys. Rev. B {\bf 61}, 1160 (2000).

\bibitem{San02d}
A.~W.~Sandvik, Phys. Rev. Lett. {\bf 89}, 177201 (2002).

\bibitem{Vaj02}
O.~P.~Vajk and M.~Greven, Phys. Rev. Lett. {\bf 89}, 177202 (2002).

\bibitem{Skn04}
R. Sknepnek, T. Vojta, and M. Vojta, Phys. Rev. Lett. {\bf 93},
097201 (2004).

\bibitem{Yu05}
Rong Yu, Tommaso Roscilde, and Stephan Haas, Phys. Rev. Lett. {\bf 94},
197204 (2005)

\bibitem{San06d}
A.~W.~Sandvik, Phys. Rev. Lett. {\bf 96}, 207201 (2006).


\bibitem{Nig92}
Nigel~Goldenfeld, {\it Lectures On Phase Transitions And The Renormalization 
Group (Frontiers in Physics)} (Addison-Wesley, 1992).

\bibitem{Cam02}
M.~Campostrini, M.~Hasenbusch, A.~Pelissetto, P.~Rossi,
and E.~Vicari, Phys.~Rev.~B {\bf 65}, 144520 (2002).

\bibitem{Pel02}
Andrea Pelissetto and Ettore Vicari, Physics Reports 368 (2002) 549-727.

\bibitem{Wan06}
L. Wang, K. S. D. Beach, and A. W. Sandvik,
Phys. Rev. B {\bf 73}, 014431 (2006).

\bibitem{Alb08}
A. F. Albuquerque, M. Troyer, J. Oitmaa, Phys. Rev. B {\bf 78},
127202 (2008).

\bibitem{Wen09}
S. Wenzel and W. Janke, Phys. Rev. B {\bf 79}, 014410 (2009).


\bibitem{Car10}
Lincoln~D.~Carr, {\it Understanding Quantum Phase Transitions (Condensed Matter Physics)} 
(CRC Press, 2010).

\bibitem{Sac11}
S.~Sachdev, {\it Quantum Phase Transitions } (Cambridge University Press, Cambridge, 2nd edition, 2011).



\bibitem{Bin81}
K. Binder, Z. Phys. B {bf 43}, 119 (1981).








\bibitem{San99}
A.~W.~Sandvik, Phys. Rev. B {\bf 66}, R14157 (1999).

\bibitem{San10}
A.~W. Sandvik, AIP Conf. Proc. 2397, 135 (AIP, New York, 2010).


\bibitem{Fri11}
L. Fritz, R. L. Doretto, S. Wessel, S. Wenzel, S. Burdin, and
M. Vojta, Phys. Rev. B {\bf 83}, 174416 (2011).



\bibitem{Har11}
Kenji Harada, Phys. Rev. E {\bf 84}, 056704 (2011).

\bibitem{Har15}
Kenji Harada, Phys. Rev. E {\bf 92}, 012106 (2015).

\bibitem{For13}
Daniel Foreman-Mackey, David W. Hogg, Dustin Lang, and Jonathan Goodman,
PUBLICATIONS OF THE ASTRONOMICAL SOCIETY OF THE PACIFIC, 125:306–312, 
2013 March.


\bibitem{Paz97}
P$\acute{a}$zm$\acute{a}$ndi,R.~T.~Scalettar, and G.~T.~ Zim$\acute{a}$nyi,
Phys. Rev. Lett. {\bf 79}, 5130 (1997)

\bibitem{Kis97}
J. Kisker and H. Rieger, Phys. Rev. B {\bf 55}, R11981 (1997).

\bibitem{Paz98}
P$\acute{a}$zm$\acute{a}$ndi and G.~T.~ Zim$\acute{a}$nyi,
Phys. Rev. B {\bf 57}, 5044 (1998).

\end{thebibliography}
\end{document}